\documentclass[12pt]{iopart}

\usepackage{algorithm}
\usepackage[noend]{algpseudocode}
\usepackage{amssymb}
\usepackage{color}
\usepackage{comment}
\usepackage{graphicx}
\usepackage{epstopdf}
\usepackage{hyperref}
\usepackage[square,sort&compress,numbers]{natbib}
\usepackage{soul}
\usepackage{tikz}
\usepackage{yfonts}
\let\epsilon\varepsilon

\newcommand\pluseq{\mathrel{+}=}
\newcommand\minuseq{\mathrel{\setminus}=}

\algnewcommand{\LineComment}[1]{\State\(\triangleright\) #1}
\algnewcommand\algorithmicinput{\textbf{Input:}}
\algnewcommand\Input{\item[\algorithmicinput]}
\algnewcommand\algorithmicoutput{\textbf{Output:}}
\algnewcommand\Output{\item[\algorithmicoutput]}

\makeatletter
\newcommand{\algmargin}{\the\ALG@thistlm}
\makeatother
\newlength{\ifwidth}
\newlength{\elseifwidth}
\algdef{SE}[parIF]{parIf}{EndparIf}[1]
  {\parbox[t]{\dimexpr\linewidth-\algmargin}{%
    \hangindent\ifwidth\strut\algorithmicif\ #1\ \algorithmicthen\strut}}{\algorithmicend\ \algorithmicif}
\algdef{C}[parIF]{parIF}{parElseIf}[1]
  {\parbox[t]{\dimexpr\linewidth-\algmargin}{%
    \hangindent\elseifwidth\strut\algorithmicelse\ \algorithmicif\ #1\ \algorithmicthen\strut}}
\algdef{Ce}[parELSE]{parIF}{parElse}{EndparIf}{\algorithmicelse}

\newlength{\forwidth}
\algdef{SE}[parFOR]{parFor}{EndparFor}[1]
  {\parbox[t]{\dimexpr\linewidth-\algmargin}{%
    \hangindent\forwidth\strut\algorithmicfor\ #1\ \algorithmicdo\strut}}{\algorithmicend\ \algorithmicfor}

\algnewcommand{\algorithmicgoto}{\textbf{go to}}
\algnewcommand{\Goto}[1]{\algorithmicgoto~\ref{#1}}

\algnewcommand{\parState}[1]{\State%
  \parbox[t]{\dimexpr\linewidth-\algmargin}{\strut #1\strut}}

\algnewcommand{\parComment}[1]{\LineComment%
  \parbox[t]{\dimexpr\linewidth-\algmargin}{\strut #1\strut}}

\makeatletter
\ifthenelse{\equal{\ALG@noend}{t}}%
{\algtext*{EndparIf}\algtext*{EndparFor}}
{}%
\makeatother

\let\oldReturn\Return
\renewcommand{\Return}{\State\oldReturn}

\let\oldfootnote\footnote
\def\footnote{\ifhmode\unskip\fi\oldfootnote}

\newcommand{\blue}[1]{{\color{blue}{#1}}}
\renewcommand{\blue}[1]{{#1}}
\let\oldst\st
\renewcommand{\st}[1]{\blue{\oldst{#1}}}

\begin{document}
\title{Inference of boundaries in causal sets}
\author{William \blue{J.} Cunningham}
\address{Department of Physics, Northeastern University, Boston, MA 02115, United States}
\ead{\href{mailto:w.cunningham@northeastern.edu}{w.cunningham@northeastern.edu}}

\begin{abstract}
\\We investigate the extrinsic geometry of causal sets in $(1+1)$-dimensional Minkowski spacetime. The properties of boundaries in an embedding space can be used not only to measure observables, but also to supplement the discrete action in the partition function via discretized Gibbons-Hawking-York boundary terms. We define several ways to represent a causal set using overlapping subsets, which then allows us to distinguish between null and non-null bounding hypersurfaces in an embedding space. We discuss algorithms to differentiate between different types of regions, consider when these distinctions are possible, and then apply the algorithms to several spacetime regions. Numerical results indicate the volumes of timelike boundaries can be measured to within $0.5\%$ accuracy for flat boundaries and within $10\%$ accuracy for highly curved boundaries for medium-sized causal sets with $N=2^{14}$ spacetime elements. 
\end{abstract}

\begin{indented}
\item[]Keywords: causal sets, quantum gravity, spacetime geometry, algorithms
\end{indented}
\submitto{\CQG}

\section{Introduction}
The causal set program~\cite{bombelli1987} is centered around the {\it \blue{Hauptvermutung}}~\cite{bombelli1989,brightwell1991,sorkin2003}, which claims that two different uniform embeddings of the same locally-finite partial order, called a causal set, into a Lorentzian manifold are nearly isometric in the Gromov-Hausdorff sense~\cite{edwards1975,gromov1981,gromov1981groups}. However, this conjecture can be understood in a much simpler way: a causal set contains all geometric and topological information about a spacetime above the discreteness scale $\ell$, up to a conformal rescaling. Though it has not yet been proven, the \blue{Hauptvermutung} is supported by many smaller results which have shown a great amount of geometric information can be extracted from a causal set, including the dimension~\cite{myrheim1978,meyer1989}, geodesic distance~\cite{brightwell1991,rideout2009}, spatial homology~\cite{major2007,major2009}, d'Alembertian~\cite{dowker2013,glaser2014,aslanbeigi2014,belenchia2016}, and Ricci curvature~\cite{benincasa2010}. \par

\begin{figure}[!t]
\centering
\includegraphics[width=0.4\linewidth]{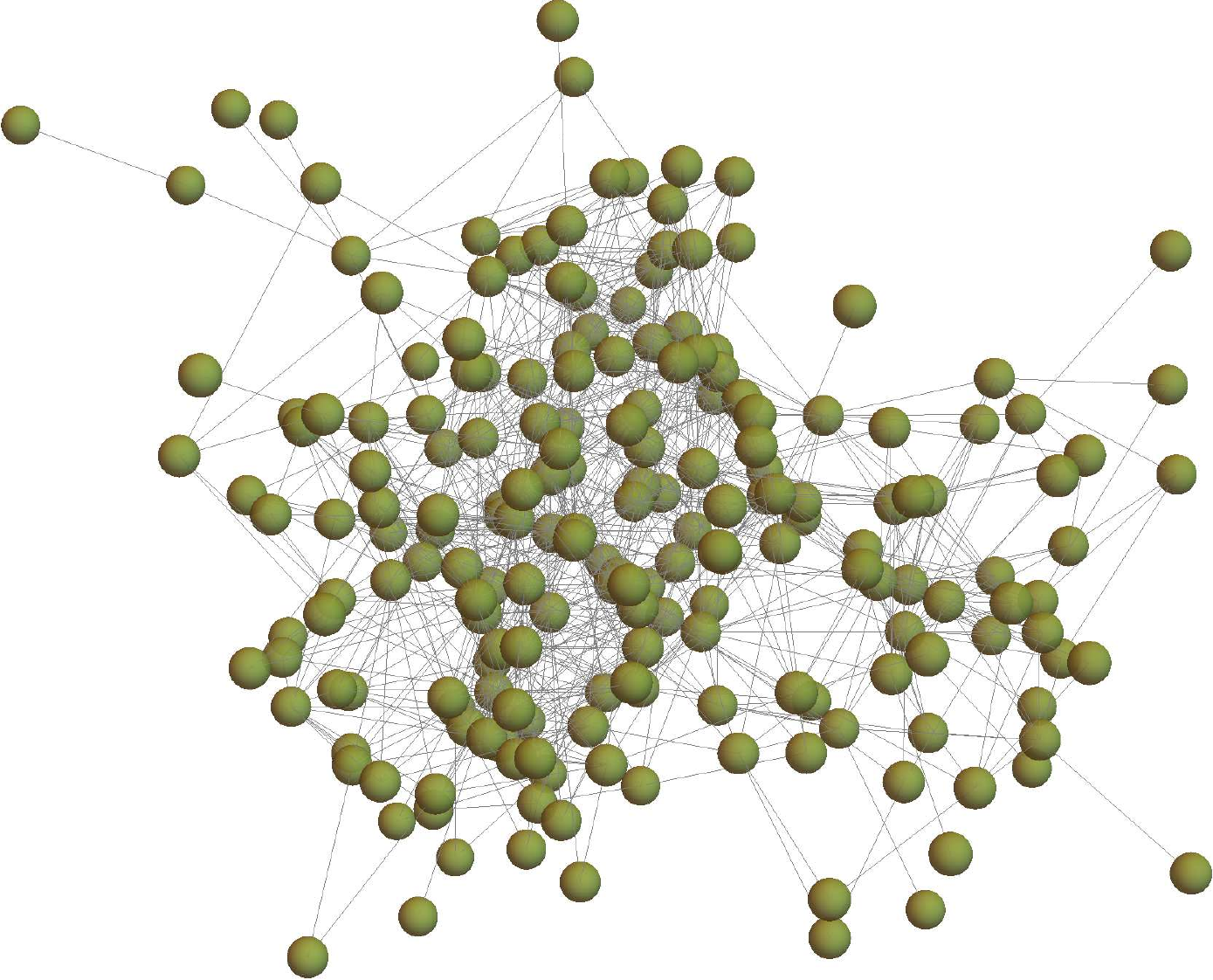}\hspace{2cm}%
\includegraphics[width=0.35\linewidth]{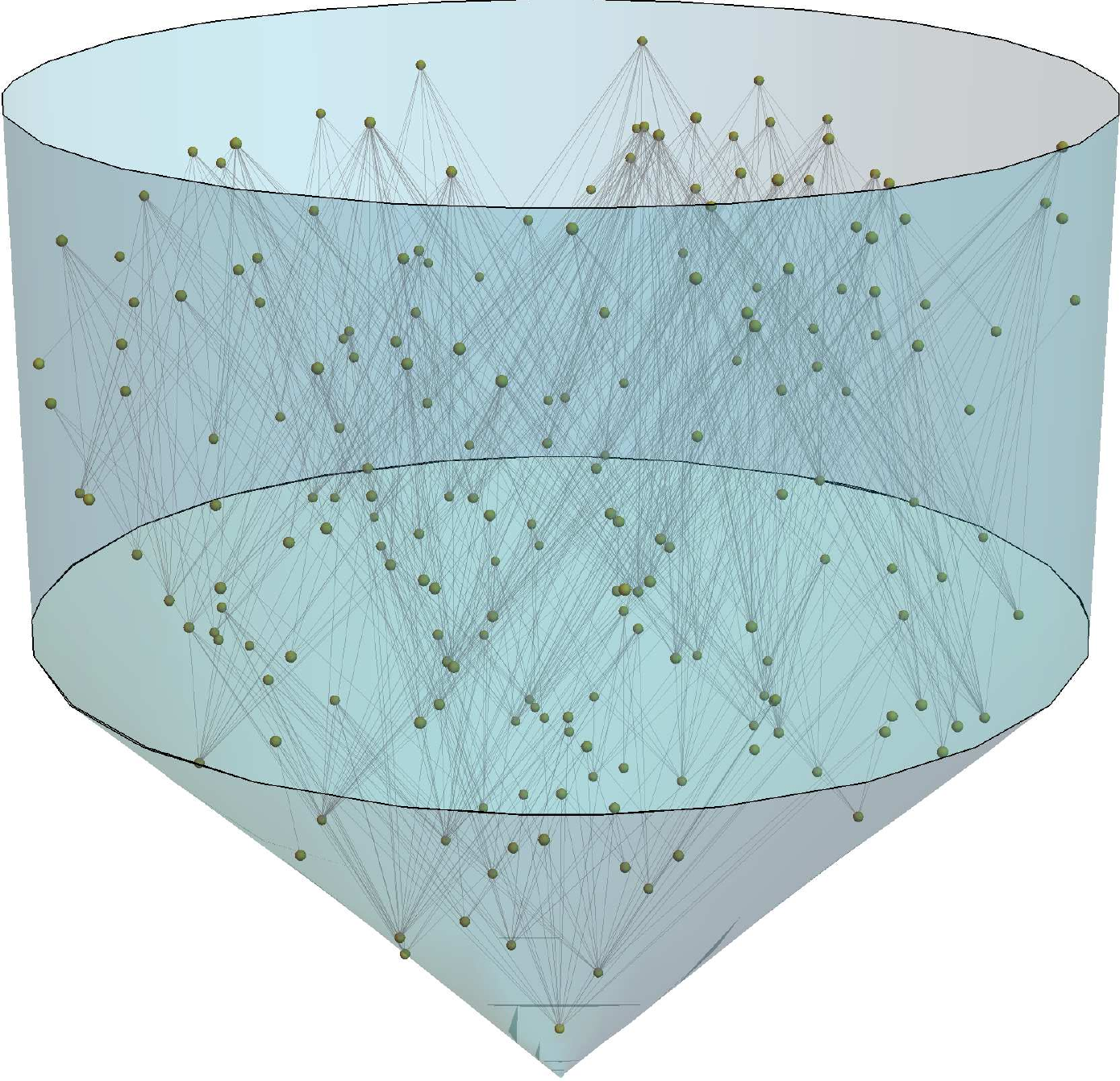}
\caption{{\bf The faithful embedding of a causal set.} An unlabeled causal set (left) with $N=200$ spacetime elements, indicated by the green points, and $5373$ causal relations, indicated by the gray lines, is faithfully embedded into the blue region (right). The causal set is bounded below by a null boundary and above by a constant-time hypersurface, with a timelike boundary of constant radius separating the two. \blue{This particular region demonstrates how in practice we can encounter causal sets with a non-trivial combination of boundaries.} A general embedding algorithm for a given causal set is unknown, but it may be possible to extract information from the causal set structure about the types of hypersurfaces which form the bounding region.}
\label{fig:embedding}
\end{figure}

Since the \blue{Hauptvermutung} describes an embedding problem \blue{(Figure~\ref{fig:embedding})}, one would hope to eventually discover an embedding method, either in the form of an analytic expression or an algorithm, to test the conjecture under certain mild assumptions \blue{(see~\mbox{\cite{clough2017}} for recent progress)}. While this is a difficult problem, one can take a first step by building a set of tools to measure extrinsic properties of causal sets with respect to an embedding space. One potential avenue has opened in the study of the Benincasa-Dowker (BD) action~\cite{benincasa2010}, i.e., the discrete analogue to the Einstein-Hilbert (EH) action,
\begin{equation}
S_{EH}=\frac{1}{2}\int_V\!R(x^\mu)\sqrt{-|g_{\mu\nu}|}\,dx^\mu\,,
\end{equation}
where $\mu,\nu\in\{0,1,2,3\}$ in four dimensions, $R(x^\mu)$ is the Ricci scalar curvature, and $g_{\mu\nu}$ is the metric tensor. Though this expression was developed during the study of intrinsic properties --- the d'Alembertian and the Ricci curvature --- it was soon noticed~\blue{\cite{benincasa2011}} it captures one of the Gibbons-Hawking-York (GHY) boundary terms~\cite{york1972,gibbons1977} which measure the contribution to the classical action from boundaries in the embedding space,
\begin{equation}
S_{GHY}=\int_{\partial V}\!K(x^i)\sqrt{|h_{ij}|}\,dx^i\,,
\end{equation}
where $i,j\in\{1,2,3\}$, $K(x^i)$ is the extrinsic curvature of the codimension-1 boundary $\partial V$, and $h_{ij}$ is the induced metric on the boundary. In the case of the {\it causal interval}, defined as the past light cone of one element intersected with the future light cone of another element, the BD action measures both the bulk term as well as the volume of the \mbox{codimension-2} surface defined by the intersection of the two light cones~\cite{benincasa2011,benincasa2013}. While it was known the BD action cannot measure the spacelike boundary terms, this observation gave hope that perhaps other boundary terms were also hidden within the expression~\cite{benincasa2011b}, which would be a good indication it held information about extrinsic geometry. Yet more recent numerical experiments have shown no other codimension-2 boundary terms are measured, and the BD action even diverges upon encountering timelike boundaries rather than recovering extrinsic geometric information~\blue{\footnote{\blue{These results will appear in a separate paper.}}}. \par

The recent discovery of the discrete boundary term for spacelike boundaries~\cite{buck2015} was a significant step forward, and it indicates the possibility of the existence of boundary terms for each type of codimension-1 and codimension-2 boundary just as in continuum physics. Yet even if there were to exist an expression akin to the spacelike boundary term for each type of boundary, it would remain unclear when such terms should be included when one calculates the full discrete action for some manifold-like causal set. This problem is compounded by the fact that in the continuum the action of a region whose boundaries approach null surfaces becomes infinite, yet the limit is finite, so there is an ambiguity over which limit any discrete structure should choose in the continuum limit. This paradox will not be studied here, but should be kept in mind. \par

In the present work we discuss several methods which allow one to infer the extrinsic geometry of boundaries in a particular causal set. We review in Section~\ref{sec:chains} the basics of chains and antichains in causal sets, and then consider two algorithms which construct and label these subsets. In Section~\ref{sec:distributions}, we discuss the properties of chains and antichains in causal sets which embed into causal intervals, with an emphasis on how one can distinguish between null boundaries and timelike or spacelike boundaries which appear to be null. We then study regions bounded by timelike hypersurfaces, and focus on two algorithms to measure the \mbox{codimension-1} volume of a timelike boundary in Section~\ref{sec:volume}. Section~\ref{sec:examples} demonstrates the performance of these numerical methods in three illustrative examples. Finally, we summarize these techniques and conclude in Section~\ref{sec:conclusion}.

\section{Review of chains and antichains}
\label{sec:chains}
\subsection{Definitions}
\label{sec:defs}
A {\it chain} is a subset in the causal set $\mathcal{C}$ which forms a clique, i.e., a total order. When a chain is inextendible with respect to the other elements in $\mathcal{C}$ it is said to be {\it maximal}, and the size of such a subset becomes a good measure of the proper time between two endpoints in the large-$N$ limit~\cite{rideout2009}. Conversely, an {\it antichain} is a subset of unrelated elements. Maximal antichains, i.e., antichains which are likewise inextendible, have been used to construct spatial hypersurfaces~\cite{major2006} and later to study spatial homology~\cite{major2009}. In this work, all chains and antichains are assumed to be maximal unless explicitly stated otherwise. We also refer to the {\it maximum} chain and antichain in a causal set, which are the largest maximal chain and antichain, respectively. \blue{While there may exist more than one maximum chain and/or antichain, the methods described in the following section are independent of the one with which we choose to work; hence, we take one of each at random in such situations.} \par

The ends of chains form the set of {\it extremal elements}: the set of {\it minimal elements} which have no past relations, $\mathcal{P}=\{x \in\mathcal{C} : \mathcal{J}^-(x)=\varnothing\}$, and the set of {\it maximal elements} which have no future relations, $\mathcal{F}=\{y\in\mathcal{C} : \mathcal{J}^+(y)=\varnothing\}$, where $\mathcal{J}^-(x)$ and $\mathcal{J}^+(y)$ respectively contain the past relations of $x$ and the future relations of $y$. One can generate a representation of the causal set consisting of chains and antichains using the following procedure. First, identify the maximum chain by measuring all possible chains with endpoints $(p\in\mathcal{P},f\in\mathcal{F})$. Then, exclude the {\it extremal pair} $(p,f)$ which bounds the maximum chain, and repeat the procedure to identify the pair which bounds the second-longest chain. This process continues until there are either no more minimal elements or no more maximal elements remaining. The set of chains which join each extremal pair is called the {\it timelike representation}. \blue{This representation is an extension of Dilworth's theorem~\cite{dilworth1950}, which allows one to partition a partially ordered set into at most $R$ chains, where $R$ is the size of the largest antichain. Here, we add the additional constraints that the chains are maximal and their extremal elements do not overlap.} \par

Whereas a chain is constructed by specifying two endpoints, an antichain is generated by providing a single seed element. The most natural method to construct antichains uses the elements of the maximum chain as the seeds. \blue{By Mirsky's theorem~\cite{mirsky1971}, a partially ordered set of height $T$ may be partitioned into $T$ antichains, and by allowing these antichains to overlap except at the seed element, we ensure each will be maximal.} This set of antichains form the {\it spacelike representation} of the causal set. Together, these two representations form the {\it spacetime representation} of the causal set. \blue{While these representations are not unique, the methods which use them remain valid, and sometimes even work better for highly symmetric causal sets.}

\subsection{Spacetime representation algorithms}
The construction of the spacetime representation deserves greater discussion for the results to be precise. The method to identify the chain length, i.e., the longest path, between a pair of causally related elements $(i,j)$ is a recursive algorithm which moves from the future to the past elements in the Alexandroff set $X_{ij}\equiv\mathcal{J}^+(i)\cap\mathcal{J}^-(j)$, recording the largest distance from each element $k\in X_{ij}$ to the final element $j$ in an array $L$ during each iteration (Algorithm~\ref{alg:chain}). The distances in $L$ are initialized to $-1$ rather than $0$ to distinguish between paths which have already been traversed and those which have not. It is possible to perform these operations efficiently if the causal set is stored in binary format and traversed using bitwise set operations~\cite{cunningham2017}. \blue{In a more complicated variation of Algorithm~\ref{alg:chain}, one may also extract the elements of the longest chain; see~\cite{cunningham2017b} for details.}\par
\begin{algorithm}[H]
\caption{Maximal Chain}
\label{alg:chain}
\begin{algorithmic}[1]
\Input
\Statex $X_{ij}$ \Comment Alexandroff set
\Statex $L$ \Comment Length array
\Statex $l$ \Comment Longest chain length
\Statex $i$ \Comment Minimal element index
\Statex $j$ \Comment Maximal element index
\Procedure{chain}{$X_{ij}$, $L$, $l$, $i$, $j$}
\For {$x\in X_{ij}$} \Comment \blue{Recursively measure length from each $x$ to $j$}
\State {$len\gets0$}
\If {$L[x]=-1$} \Comment \blue{The distance $L_{xj}$ has not been calculated}
\State {$M_{xj}\gets\mathcal{J}^+(x)\cap\mathcal{J}^-(j)$} \Comment \blue{Look at elements between $x$ and $j$}
\If {$|M_{xj}|>0$} \Comment \blue{If the Alexandroff set is not empty}
\State {$\blue{\chi}\!\gets\Call{chain}{M_{xj}, L, l, x, j}$} \Comment \blue{Find the longest distance}
\State {$L[x]\gets \blue{\chi}$} \Comment \blue{And record the results}
\State {$len\gets \blue{\chi}$}
\Else \Comment \blue{Otherwise, the distance is 1}
\State {$L[x]\gets1$}
\State {$len\gets1$}
\EndIf
\Else \Comment \blue{If it's already calculated, use the recorded value}
\State {$len\gets L[x]$}
\EndIf
\State {$l\gets$ max($len,l$)} \Comment \blue{Record the largest length}
\EndFor
\Return $l+1$
\EndProcedure
\Output
\Statex $l$ \Comment Length of the longest chain
\end{algorithmic}
\end{algorithm}
The antichain construction algorithm uses a slightly different procedure: it is a variation of the \blue{maximal independent set problem for transitively closed directed acyclic graphs~\cite{gavril1987}}. We first specify an initial seed element $x$, and then consider all other elements \blue{in the causal set $\mathcal{C}$} unrelated to $x$, \blue{$\Xi=\mathcal{C}\setminus(\mathcal{J}^+(x)\cup\mathcal{J}^-(x))$}, as potential candidates for the antichain $\mathcal{A}$, \blue{so that initially $\mathcal{A}_\blue{\Xi}=\{x\cup\Xi\}$ and by the end of the procedure $|\Xi|\to0$ and $\mathcal{A}_\Xi\to\mathcal{A}$}. In each iteration of the algorithm, for each \blue{$y\in\Xi$} we measure the number of elements $c$ \blue{which would remain in $\mathcal{A}_\blue{\Xi}$ if the relations of $y$ were removed}, i.e., $c=|\blue{\mathcal{A}_\Xi\setminus(}\,\mathcal{J}^-(y)\cup\mathcal{J}^+(y)\blue{)}|$. \blue{Keeping} the element $y\,\,\blue{\in\mathcal{A}_\Xi}$ which \blue{maximizes} $c$ maximizes the size of the final antichain \blue{$\mathcal{A}$}. This procedure continues until no candidates remain, \blue{at which point $\mathcal{A}_\Xi\to\mathcal{A}$ is a true maximal antichain}. The details of this algorithm are provided in Algorithm~\ref{alg:antichain}. \blue{It is important to note that Algorithm~\mbox{\ref{alg:antichain}} is a \textit{greedy} algorithm, meaning it uses a short-term optimization to avoid considering all possible antichains. While it will only very rarely produce the true maximum antichain, it is still useful to measure width, since the true width is directly proportional to that given by this algorithm. The method could easily be modified to a non-greedy version by taking all possible $y\in\Xi$ at each step, and using a recursive method as in Algorithm~\ref{alg:chain}.} These particular algorithms \blue{as well as the others described in this work} are implemented in C and x86 Assembly with OpenMP and AVX optimization as part of the {\it Causal Set Generator}~\cite{cunningham2017b}.
\begin{algorithm}[t]
\caption{Maximal Antichain}
\label{alg:antichain}
\begin{algorithmic}[1]
\Input
\Statex $x$ \Comment Antichain seed
\Statex $\Xi$ \Comment Antichain candidates
\Procedure{antichain}{$x$, $\Xi$}
\State {$\mathcal{A}_\blue{\Xi}\gets \{x\cup\Xi\}$} \Comment \blue{Initially consider $x$ and elements unrelated to $x$}
\While {$|\Xi|>0$} \Comment \blue{Continue until no candidates remain}
\State $\blue{\sigma}\gets0$, $z\gets0$
\For {$y\in\blue{\Xi}$} \Comment \blue{Consider each candidate}
\State $\blue{c}\gets|\mathcal{A}_\blue{\Xi}-(\mathcal{J}^-(y)\cup\mathcal{J}^+(y))|$ \Comment \blue{Find how many elements remain}
\If {$\blue{c}>\blue{\sigma}$} \Comment \blue{Record the element which maximizes $c$}
\State $\blue{\sigma}\gets \blue{c}$
\State $z\gets y$
\EndIf
\EndFor
\State $\mathcal{A}_\blue{\Xi}\minuseq\mathcal{J}^-(z)\cup\mathcal{J}^+(z)$ \Comment \blue{Remove the neighbors $\mathcal{A}_\Xi$}
\State $\Xi\minuseq\mathcal{J}^-(z)\cup\mathcal{J}^+(z)\cup z$ \Comment \blue{Remove neighbors plus the element from $\Xi$}
\EndWhile
\State \blue{$\mathcal{A}\gets\mathcal{A}_\Xi$} \Comment \blue{When complete, $\mathcal{A}_\Xi$ will be the antichain}
\Return $\mathcal{A}$
\EndProcedure
\Output
\Statex $\mathcal{A}$ \Comment A maximal antichain
\end{algorithmic}
\end{algorithm}

\subsection{Assumptions}
\label{sec:assumptions}
The spacetime representation admits a natural scheme for ordering chains and antichains. Antichains are labeled according to the distance of the seed element from the minimal element in the maximum chain, i.e., they are time-ordered with respect to the seed element. For this to remain valid, we must assume the causal set embeds into a conformally flat manifold. The chain ordering is performed by ranking the elements which intersect with the maximal antichain by spatial distance from the {\it representation-induced origin} of the causal set, defined as the element at the intersection of the maximum chain and maximum antichain. The algorithm to determine this inferred spatial distance is discussed in more detail in Section~\ref{sec:volume}. \par

The causal sets used in numerical experiments are generated by Poisson sprinkling $N$ elements into a fixed-volume region, where $N$ is a Poisson random variable with a specified mean $\bar{N}$. When reference is made to expectations of observables and averages over graphs, we assume the causal sets we study are generated from an ensemble which reliably provides causal sets with the same boundary geometry in the thermodynamic limit $\bar{N}\to\infty$, though for any single large causal set one can argue its observables take values close to the mean. \par

Finally, for the following results to hold, we assume the size of the maximum chain is large, $T\gtrsim2^{6}$, the size of the maximum antichain is large, $R\gtrsim2^{6}$, and the causal set is relatively large, $N\gtrsim2^{10}$. Furthermore, when the inverse extrinsic curvature $K^{-1}$ of a non-null boundary is on the order of the discreteness scale $\ell$ of the causal set, the boundary is indistinguishable from a null boundary (Section~\ref{sec:distributions}), so in the following measurement of the boundary volume (Section~\ref{sec:volume}) we assume any non-null boundary is smooth, continuous, slowly varying, and has an inverse extrinsic curvature much larger than the discreteness scale, $\ell K\ll1$. 

\section{Characteristics of the causal interval}
\label{sec:distributions}
\subsection{Chain and antichain profiles}
\begin{figure*}[!t]
\centering
\vspace{-1cm}%
\includegraphics[width=\linewidth]{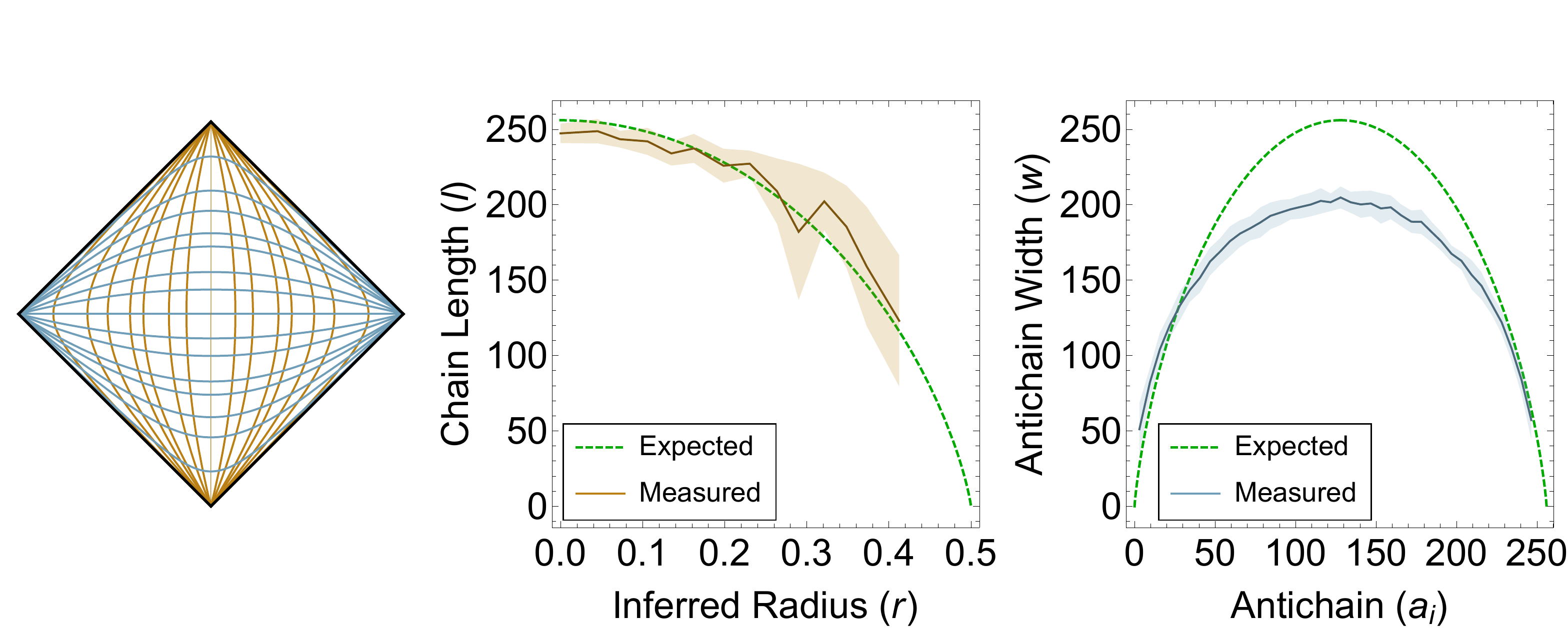}
\caption{{\bf The spacetime representation for the causal interval.} The representation of the causal interval in the $(1+1)$-dimensional Minkowski spacetime is shown in the left panel, where the orange vertical curves correspond to the expected paths of chains and the blue horizontal curves correspond to the expected paths of antichains. The orange and blue straight lines crossing the center, \blue{denoted the representation-induced origin,} respectively represent the maximum chain and antichain. In the center panel, the empirical chain lengths (orange) fall nearly perfectly across the expected values (green), given by~(\ref{eq:geo_to_chain}). The radial coordinates are inferred by averaging over values sampled from the marginal distribution~(\ref{eq:marginal_x}).  Fluctuations increase with radial distance due to finite-size effects. The right panel shows the antichain widths, i.e., cardinalities, for the same causal sets, ranked by a time coordinate inferred from the intersection of each antichain $a_i$ with the maximum chain. All data is averaged over ten graphs with height $L=1$ and size $N=2^{14}$, and the shaded regions indicate the standard deviation of the mean.}
\label{fig:null_profiles}
\end{figure*}

The first challenge in characterizing a timelike or spacelike boundary is distinguishing it from a null boundary. We can characterize the ordered sets of chain and antichain sizes, hereafter called {\it profiles}, for an interval of height $L$ in $d$-dimensional Minkowski spacetime using the spacetime representation described in the previous section. The continuum limit of the chain representation can be modeled by the family of hyperbolic curves which pass through the bottom of the interval, $T_p$, the top of the interval $T_f$, and some point $(0,r)$, shown by the orange curves in Figure~\ref{fig:null_profiles}(left). The continuum length of a chain passing through the waist ($t=0$) at radius $r$ is
\begin{equation}
l(r)=\frac{1}{2}\sqrt{4\zeta(r)^2-L^2}\ln\left(\frac{4\zeta(r)}{2\zeta(r)-L}-1\right)\,,
\end{equation}
where $\zeta(r)\equiv r/2+L^2/(8r)$. \par

The continuum length $l(r)$ is directly proportional to the discrete graph distance $D(r)$~\cite{brightwell1991}. For instance, in $(1+1)$-dimensional Minkowski spacetime, 
\begin{equation}
\label{eq:geo_to_chain}
\mathbb{E}\left[D(r)\right]=\sqrt{2}\,l(r)/\ell\,,
\end{equation}
\blue{where $\mathbb{E}[x]$ refers to the expectation value of some variable $x$.} Since the spatial distribution of chains is not uniform, we approximate the spatial distribution $\rho(x)$ for maximal elements $\mathcal{F}$ by considering a Poisson point process inside the region between the future null boundary and the hyperbolic surface at proper time $\ell$ to the past of the boundary. This gives a marginal distribution
\begin{equation}
\label{eq:marginal_x}
\rho(x) = \left(1/2-x-\zeta(x)+\sqrt{x^2+\zeta(x)^2-L^2/4}\right)/\tilde{V}\,,
\end{equation}
where $\tilde V$ is the volume of the region described. Hence, when comparing measured chain lengths to the theoretical profile for a known region, one should sample $x$ from this distribution. \par

By symmetry, the same arguments can be used to calculate the width of an antichain centered about the origin. The width $w(t)$ of an antichain passing through $r=0$ at time $t$ is equal to the length of a chain passing through $t=0$ at spatial distance $|t|$, multiplied by half the volume of the $(d-2)$-sphere $S_{d-2}$,
\begin{equation}
w(t)=l(|t|)S_{d-2}/2\,,
\end{equation}
where $S_d=(d+1)\pi^{(d+1)/2}/\Gamma((d+1)/2+1)$. The antichain width is translated to the discrete setting in the same way as the chain length:
\begin{equation}
\label{eq:expected_antichain}
\mathbb{E}\left[W(t)\right] = l(|t|)S_{d-2}/(\sqrt{2}\ell)\,,
\end{equation}
where $W(t)$ is the discrete antichain width. Using these expressions, the chain and antichain profiles are shown in Figure~\ref{fig:null_profiles}(center,right). \blue{While the antichain width measured by Algorithm~\mbox{\ref{alg:antichain}} does not exactly match that given by~(\mbox{\ref{eq:expected_antichain}}), the functional form is the same, making it a good enough measure of width for the purposes of the following experiments. It is believed that the ratio of the peaks is a constant dependent only on dimension, which we leave as an open problem for future study.} \par

\begin{figure*}[!t]
\centering
\includegraphics[width=\linewidth]{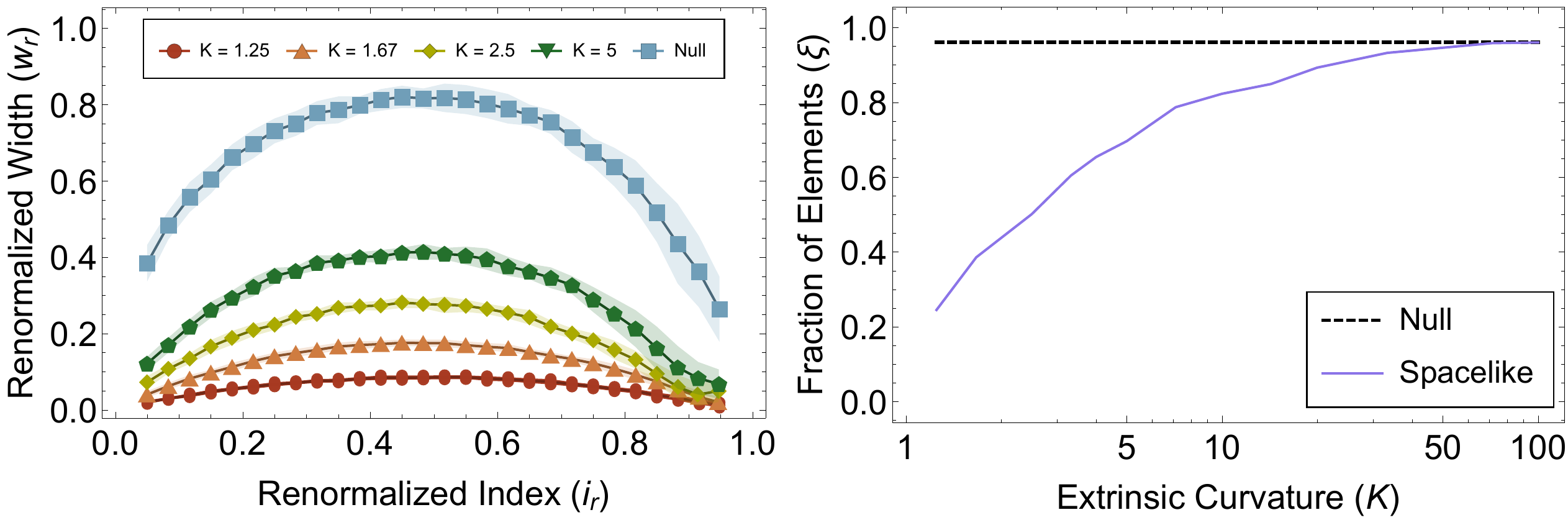}
\caption{{\bf Timelike and spacelike boundaries.} The left panel compares the renormalized antichain widths, $w_r=w/T$, for causal sets in several regions bounded by constant-curvature timelike hypersurfaces (red, orange, yellow, green) to those of causal sets with a null boundary (blue). The values are renormalized to lie in the range $[0,1]$ to account for changes in volume of different regions. The right panel shows the fraction of elements $\xi\equiv |\blue{X_{pf}}|/N$ which lie in the Alexandroff set $\blue{X_{pf}}$ defined by the extremal pair $(p,f)$ of the maximum chain for causal sets in regions bounded by spacelike hypersurfaces with variable extrinsic curvature $K$ (purple). In both cases, the boundaries are indistinguishable from null ones when $\ell K\to1$. All data is averaged over ten causal sets of size $N=2^{14}$, and the shaded regions indicate the standard deviation of the mean.}
\label{fig:profiles}
\end{figure*}

\subsection{Comparison of timelike and null boundaries}
Using the two profiles in Figure~\ref{fig:null_profiles} for reference, one can compare causal sets from a region with timelike boundaries to those from one with a null side. In general, the chain profile is used to detect the top and bottom corners of the interval, and the antichain profile to detect the side corners. Therefore, to study timelike boundaries we focus on the antichain profile in particular. By studying a family of causal sets bounded by constant-curvature timelike surfaces one can show their antichain profiles converge toward the profile for the null boundary as $\ell K\to1$ (Figure~\ref{fig:profiles}(left)). The {\it renormalized index} $i_r\equiv a_i/T$ is simply the antichain index $a_i$ rescaled by the length of the maximum chain $T$, and the renormalized width $w_r=w/T$ likewise is the rescaled size of each antichain. 

\subsection{Comparison of spacelike and null boundaries}
One method to characterize spacelike boundaries is to examine the chain profile, but Figure~\ref{fig:null_profiles} indicates not very many chains are selected compared to the number of antichains, and the fluctuation in lengths tends to increase for chains with $r\sim r_{max}$. Another way to characterize the boundary is to consider the size of the Alexandroff set $\blue{X_{pf}}$ of the extremal pair $(p,f)$ of the maximum chain. For causal sets embedded in a causal interval, $|\blue{X_{pf}}|$ converges to the size of the entire causal set, whereas in a region with spacelike boundaries it does not. The right panel of Figure~\ref{fig:profiles} shows the fractional cardinality $\xi\equiv|\blue{X_{pf}}|/N$ for causal sets in regions with constant-curvature spacelike boundaries as well as for those in the causal interval. \par

\section{The boundary volume}
\label{sec:volume}
Once we have distinguished the types of boundaries of an embedded causal set, we can confidently measure their volumes and other properties. In the following analysis, we assert $\ell K\ll1$, as mentioned in Section~\ref{sec:assumptions}, to avoid further discussion about ambiguities. We begin by reviewing the analytic expression for the volume of spacelike boundaries, and then discuss algorithms for measuring the volume of timelike boundaries. In $(1+1)$-dimensional Minkowski spacetime, \mbox{codimension-2} corners enter as $0$-dimensional points, so the following discussion only covers numerical methods for their identification.

\subsection{Review of spacelike boundaries}
The volume of spacelike boundaries in causal sets was first reported in~\cite{buck2015}. Given the number of minimal elements $P_0$ and maximal elements $F_0$ one may write the volumes of the past and future boundaries, $\Sigma^-$ and $\Sigma^+$ respectively, as
\begin{eqnarray}
V_{\Sigma^-} &=& \left(\frac{\ell}{l_p}\right)^{d-1}\frac{b_d}{\Gamma(1/d)}F_0\,, \\
V_{\Sigma^+} &=& \left(\frac{\ell}{l_p}\right)^{d-1}\frac{b_d}{\Gamma(1/d)}P_0\,,
\end{eqnarray}
where
\begin{equation}
b_d = d\left(\frac{S_{d-2}}{d(d-1)}\right)^{1/d}\,,
\end{equation}
$l_p$ is the Planck length, and $S_{d-2}$ is the volume of the $(d-2)$-sphere. In practice, the continuum volume is compared to $l_p^{d-1}V_{\Sigma^\pm}$. The convergence of these expressions is studied in Section~\ref{sec:examples}.

\subsection{Timelike boundaries}
While it is easy to identify and measure spacelike boundaries in a causal set, it is challenging to do the same for timelike boundaries. The following procedure solves this problem by first detecting these elements and then building chains which cover the boundary.

\subsubsection{Boundary element detection}
Unlike the set of extremal elements which cover a spacelike boundary, the subset of elements $\mathcal{T}$ in the causal set $\mathcal{C}$ which cover a timelike boundary is not trivial to quickly pick out of a causal set. We can distinguish these elements from the {\it internal elements} located deep in the bulk first by observing that they have far fewer relations in expectation. In a faithful embedding into curved spacetime, the number of relations of elements along the timelike boundary also varies in the temporal direction. Therefore, we only compare elements along a spatial hypersurface, or antichain. This is not one of the maximal antichains described previously, but rather one of the set partitions generated when the causal set is partitioned into antichains. \par

\begin{figure*}[!t]
\centering
\includegraphics[width=\linewidth]{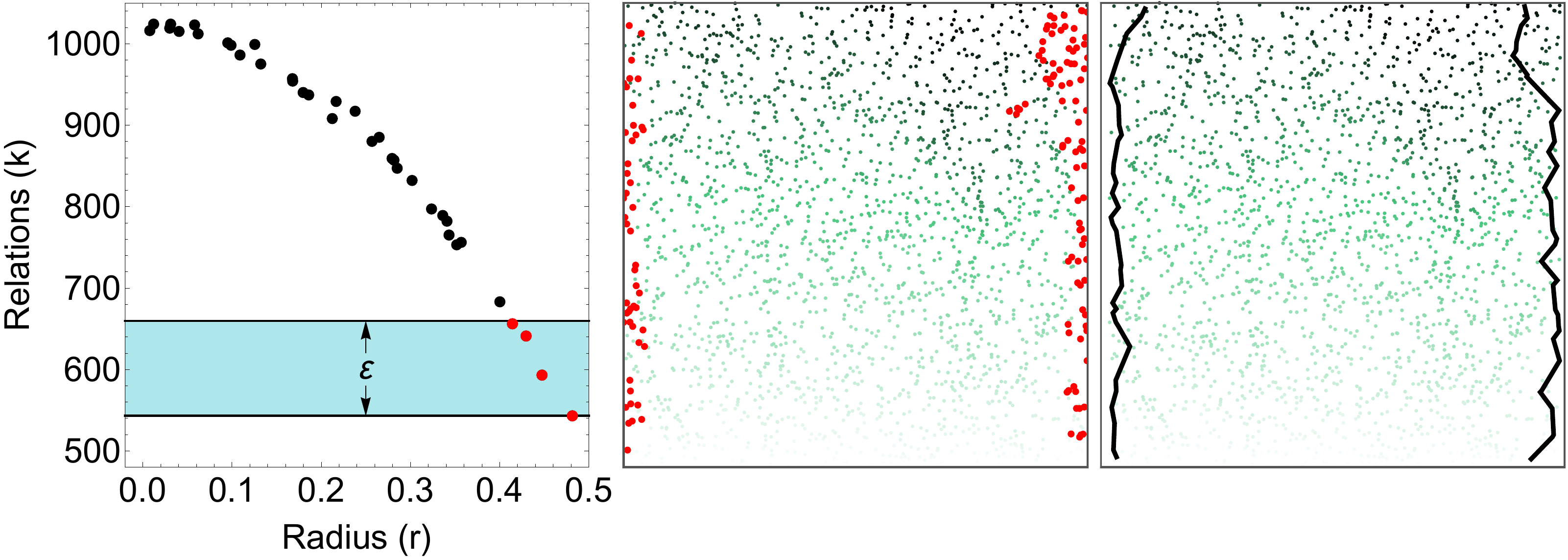}
\caption{{\bf Measurement of timelike boundary volume.} The causal set is partitioned into antichains, each of whose elements lie at a constant graph distance to the minimal elements $\mathcal{P}$. In each antichain, elements near the timelike boundary have the fewest number of relations (left). Those with a number of relations in the range $k\in[k_{min},k_{min}+\epsilon)$ are selected as candidates (red). The center panel shows the resulting set of candidates $\mathcal{T}$ on top of the antichain partitions, where each partition's elements are the same shade of green. Maximal chains $\{\mathcal{B}_i\}$, which are proxies for timelike geodesics, are then constructed by maximizing the number of elements $\mathcal{T}$ in each chain, shown by the bold black lines (right). \blue{The origin is always taken to be at the center of the region to enable a natural extension to higher dimensions.}}
\label{fig:timelike}
\end{figure*}

The antichains are constructed by assigning to each element the maximum graph distance from \blue{that} element to the minimal elements, \blue{i.e., for element $n$ the distance is $t_n=$ max(\textproc{chain}$(p, n)$) for all minimal elements $p\in\mathcal{P} : p \prec n$, where \textproc{chain}$(p,n)$ indicates the length of the longest chain between elements $p$ and $n$. Hence, each antichain is defined by the set of elements with equal $t_n$}. The correlation between the number of relations and spatial distance from the origin is shown for the causal set embedded into a square in Figure~\ref{fig:timelike}(left). The elements with a number of relations in the range $k\in[k_{min},k_{min}+\epsilon)$ are selected from each antichain as potential candidates to cover the timelike boundary, shown in the center panel of Figure~\ref{fig:timelike}. The depth $\epsilon$ adjusts the algorithm to select elements within a variable spatial distance from the boundary. The algorithm which selects a causal subset $\mathcal{T}\subset\mathcal{C}$ is shown in Algorithm~\ref{alg:candidates}.\par
\begin{algorithm}
\caption{Boundary Candidates}
\label{alg:candidates}
\begin{algorithmic}[1]
\Input
\Statex $\mathcal{C}$ \Comment The causal set
\Statex $\mathcal{P}$ \Comment Minimal elements
\Statex $k$ \Comment Number of relations per element
\Statex $\epsilon$ \Comment Boundary depth

\Procedure{chain}{$i$, $j$} \Comment \blue{This is a helper function for the procedure below}
\State $X_{ij}\gets\mathcal{J}^+(i)\cap\mathcal{J}^-(j)$
\State $L\gets\{\}$ \Comment \blue{Empty array}
\Return $\Call{chain}{X_{ij},L,0,i,j}$ \Comment \blue{The longest chain between $i$ and $j$ in $\mathcal{C}$}
\EndProcedure

\Procedure{candidates}{$\mathcal{C}$, $\mathcal{P}$, $k$, $\epsilon$}
\State $\mathcal{T}\gets\{\}\,,t_\blue{n}\blue{\gets -1\,\forall\,n}$
\For {$p\in\mathcal{P}$ and $n\notin\mathcal{P}$} \Comment \blue{$p$ is a minimal element; $n$ is not}
\If {\blue{$p\nprec n$}}
\State \blue{\textbf{continue}}
\EndIf
\State $t_\blue{n}\gets$ max($t_\blue{n}, \Call{chain}{p,n}$) \Comment \blue{Record the longest distance from $t_n$ to the $p$'s}
\EndFor
\State $\kappa\gets\{\infty,\ldots,\infty\}$
\For {$i=\{0,\ldots,\mathrm{max}(t)-1\}$} \Comment \blue{In each of the antichain partitions$\ldots$}
\For {$n\in\mathcal{C}$} \Comment \blue{Record the fewest relations}
\If {$t_\blue{n} = i$}
\State $\kappa[i]\gets$ min($\kappa[i],k[n]$)
\EndIf
\EndFor
\For {$n\in\mathcal{C}$} \Comment \blue{Record elements with few relations}
\If {$t_\blue{n} = i$ {\bf and} $k[n] < \kappa[i]+\epsilon$} \Comment \blue{i.e., within the minimum plus $\epsilon$}
\State $\mathcal{T}$.append($n$)
\EndIf
\EndFor
\EndFor
\EndProcedure
\Output
\Statex $\mathcal{T}$ \Comment The candidate elements
\end{algorithmic}
\end{algorithm}

\subsubsection{Timelike Boundary Measurement}
The second part of the procedure uses the candidate elements $\mathcal{T}$ to build chains $\{\mathcal{B}_i\}$ which cover the timelike boundary. The method is similar to the one described in Section~\ref{sec:defs} which formed the set of extremal pairs. Using the maximal and minimal elements within the subset $\mathcal{T}$, maximal chains $\{\mathcal{B}_i(\mathcal{T})\}$ are formed using only the candidates $\mathcal{T}$. For each adjacent pair of elements in a chain, i.e., $\{(x,y)\in\mathcal{B}_i(\mathcal{T}) : \blue{X_{xy}}=\varnothing\}$, a maximal chain is constructed between $x$ and $y$ using the elements \blue{$\{z\in\mathcal{C}\setminus\mathcal{T}\}$}. This guides the chain along the boundary, enabling us to measure the boundary by incorporating elements in the full causal set rather than just the candidate elements. The longest chain $\mathcal{B}_{max}$ is taken to be a good cover of the boundary in a particular region, and then the elements which form that chain are removed from $\mathcal{T}$. \par

In $(1+1)$-dimensions, only the two longest chains are taken to cover the timelike boundaries, but in higher dimensions the procedure is repeated while $|\mathcal{B}_{max}|>\delta$ for some $\delta$. In practice, if the procedure continues until $\mathcal{T}=\varnothing$, one can see a sharp drop in the {\it chain weight}, defined as the number of elements in $\mathcal{T}$ occurring in $\mathcal{B}_i$, and this transition can be used to pick $\delta$. Those chains with size smaller than $\delta$ typically cover regions already covered by longer chains. The algorithm describing this procedure is given in Algorithm~\ref{alg:timelike_measurement} and the result is shown in Figure~\ref{fig:timelike}(right). Once the total number of elements $\tau=\sum_i|\mathcal{B}_i|$ in the set of chains has been measured, the continuum length may be recovered via~(\ref{eq:geo_to_chain}). \par
\begin{algorithm}
\caption{Boundary Measurement}
\label{alg:timelike_measurement}
\begin{algorithmic}[1]
\Input
\Statex $\mathcal{C}$ \Comment The causal set
\Statex $\mathcal{T}$ \Comment Boundary candidates
\Statex $\delta$ \Comment Chain length threshold
\Procedure{ax\_set}{$\mathcal{S}$, $i$, $j$} \Comment \blue{This is a helper function for the procedure below}
\Return $\mathcal{J}^+_\mathcal{S}(i)\cap\mathcal{J}^-_\mathcal{S}(j)$ \Comment \blue{The Alexandroff set using elements $i$, $j$ in set $\mathcal{S}$}
\EndProcedure

\Procedure{timelike\_volume}{$\mathcal{C}$,$\mathcal{T}$,$\delta$}
\State $\mathcal{P}\gets \{t\in\mathcal{T} : \mathcal{J}^-_\mathcal{T}(t)=\varnothing\}$ \Comment \blue{The minimal boundary elements} \label{op:first}
\State $\mathcal{F}\gets \{t\in\mathcal{T} : \mathcal{J}^+_\mathcal{T}(t)=\varnothing\}$ \Comment \blue{The maximal boundary elements}
\State $L\gets\{\},\,\mathcal{B}_{max}\gets\{\},\,l_{max}\gets0$
\For {$p\in\mathcal{P}$ {\bf and }$f\in\mathcal{F}$} \Comment \blue{For all pairs of minimal/maximal elements}
\State $\blue{X_{pf}}\gets$ \Call{ax\_set}{$\mathcal{T},p,f$} \Comment \blue{The Alexandroff set using only $\mathcal{T}$}
\State $\{\blue{l_{pf}},\mathcal{B}_X\}\gets\Call{chain}{\blue{X_{pf}},L,0,p,f}$ \Comment \blue{The longest chain $\mathcal{B}_X$ and its length}
\For {$(m, n)\in\mathcal{B}_X :$ \Call{ax\_set}{$\mathcal{T},m,n$} $= \varnothing$} \Comment \blue{For each link in $\mathcal{B}_X$}
\State $\blue{X_{mn}}\gets$ \Call{ax\_set}{$\mathcal{C},m,n$} \Comment \blue{Find the longest chain using $\mathcal{C}$}
\State $\blue{l_{pf}}\pluseq\Call{chain}{\blue{X_{mn}},L,0,m,n}$
\EndFor
\If {$\blue{l_{pf}} > l_{max}$} \Comment \blue{Record the longest chains and lengths}
\State $l_{max}\gets \blue{l_{pf}}$
\State $\mathcal{B}_{max}\gets\mathcal{B}_X$
\EndIf
\EndFor
\If {$l_{max}>\delta$} \Comment \blue{If the chain is long enough, it is a good cover}
\State $\tau\pluseq l_{max}$
\State $\mathcal{T}\minuseq\mathcal{B}_{max}$
\State \Goto{op:first} \Comment \blue{Continue until no good covers remain}
\EndIf
\EndProcedure

\Output
\Statex $\tau$ \Comment The boundary volume
\end{algorithmic}
\end{algorithm}

\subsubsection{Convergence}
To demonstrate convergence, we claim in the $N\to\infty$ limit the chains $\{\mathcal{B}_i\}$ will perfectly cover the timelike boundaries. This can only occur if $\epsilon$ is controlled in a way that the number of elements in $\mathcal{T}$ grows like the \mbox{codimension-1} volume of the timelike boundary, in units of $\ell$, rather than the like number of elements $N$. Hence, if $\epsilon$ is chosen such that the number of candidate elements per antichain grows like $N^{d-2}$, and $\epsilon$ is as small as possible such that the causal subset \blue{$\mathcal{C}_\mathcal{T}$} defined by the elements of $\mathcal{T}$ is percolated, \blue{i.e., $\mathcal{C}_\mathcal{T}$ has two connected components in $(1+1)$ dimensions or one connected component in higher dimensions,} then the elements of $\mathcal{T}$ will always remain close to the timelike boundary. Since it is known maximal chains converge to timelike geodesics as $N\to\infty$~\cite{brightwell1991}, then the measured boundary volume will converge to the continuum volume when $\epsilon$ is bounded using this prescription.

\subsection{Corners}
\label{sec:corners}
The codimension-2 boundaries are known as corners, and they arise due to the intersections of codimension-1 boundaries. Detecting corners induced by spacelike-timelike boundary intersections is easy, since the corner elements are simply the extremal elements of the chains covering the timelike boundaries, i.e.,  $(\mathcal{P}\cap\mathcal{B})\cup(\mathcal{F}\cap\mathcal{B})$. When a corner has an obtuse angle, it can be difficult to infer its presence from the chain and antichain profiles alone. It can be helpful to use another profile as well, called the {\it Alexandroff profile}, to characterize the hypersurface. One measures the size of the Alexandroff set \blue{$X_{pf}$} defined by each of the chain's extremal pairs \blue{$(p,f)$}, and watches how its size changes with the inferred radial distance. To detect these corners, the graph density must be very large to get an accurate measurement of the derivatives of the Alexandroff and chain profiles for small renormalized index. If they never tend to zero, we can remain confident a corner actually exists. \par

\begin{figure}[!t]
\centering
\includegraphics[width=\linewidth]{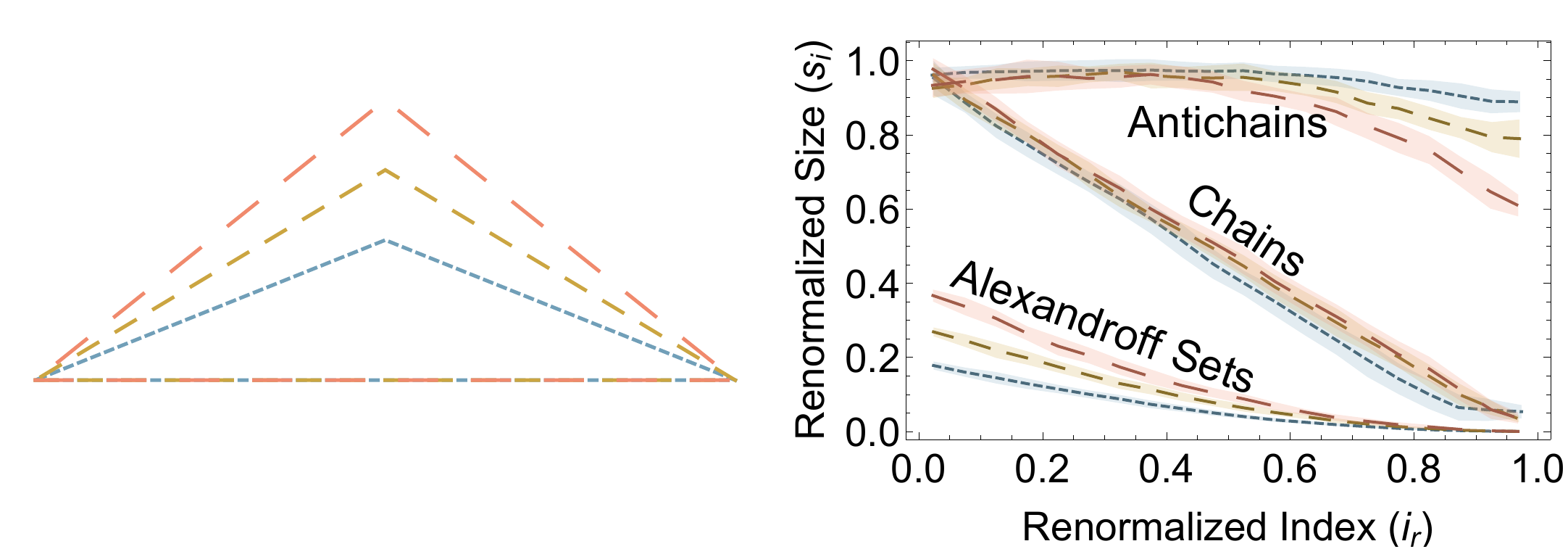}
\caption{{\bf Detection of codimension-2 corners.} Causal sets embedded into different triangular regions (left) have codimension-2 corners. The corner formed by two spacelike hypersurfaces intersecting at an acute angle is characterized by a large difference in the chain and antichain sizes at large renormalized index (right). In particular, the antichain size never decreases toward zero unless one of the hypersurfaces approaches the null limit. The renormalized size is equivalent to the renormalized width $w_r$ for antichains, the renormalized length $l_r$ for chains, and fractional Alexandroff set size $\xi$ for Alexandroff sets. Data are averaged over ten causal sets of size $N=2^{13}$, and the shaded regions indicate the standard deviation of the mean.}
\label{fig:corners}
\end{figure}
When the corner's angle is acute, it is somewhat easier to determine its presence. Figure~\ref{fig:corners} demonstrates what the chain, antichain, and Alexandroff profiles look like for an isosceles triangle defined by the points $\{(0,-1),(0,1),(t_0,0)\}$. The renormalized size $s_i$ for chains and antichains refers to the renormalized length and width, respectively, whereas for the Alexandroff profile $s_i=|X_{ij}|/N$. The acute angle is characterized by the large difference in the two profiles: chains whose lengths go to zero at large radius combined with antichains which are always large indicate there is no timelike or null boundary. While it may appear the slope of the chain profile in Figure~\ref{fig:corners} could measure the angle, preliminary experiments indicate neither the slope of the chain profile nor that of the Alexandroff profile are reliable metrics.

\section{Examples}
\label{sec:examples}
Finally, we consider several examples in various regions of $(1+1)$-dimensional Minkowski spacetime. In each case, we look at how the chain, antichain, and Alexandroff profiles can be used together to identify the shape of a bounding region in a flat embedding space and estimate the boundary volume. \blue{It is important to emphasize that the following arguments are useful as a first step towards characterizing the boundary, and in practice it is best to compare results to the profiles of causal sets with known boundaries which are generated from sprinklings~\mbox{\cite{cunningham2017}}.} \par

Each example highlights a certain difficulty or ambiguity which one might encounter in practice. When we measure timelike boundaries, we take the smallest $\epsilon$ such that at least two elements are selected from each antichain partition, and we take the largest two chains in $\{\mathcal{B}_i\}$. All data shown is averaged over ten graphs of size $N=2^{11}$ unless otherwise indicated. \par

\subsection{The square and the cylinder}
\label{sec:sq_cyl}
The first example demonstrates how one might differentiate between a causal set in a square region with flat timelike boundaries and one in a region with no spatial boundaries, i.e., the surface of a 2-cylinder. The chain, antichain, and Alexandroff profiles are shown for the causal sets in the square in Figure~\ref{fig:sq_v_cyl}(left). The chain and antichain profiles remain nearly constant, indicating the boundary shape is likely flat and symmetric. The chain profile always decreases slightly even when the spacelike boundaries are flat and constant, since the chain distribution can never be uniform when there are Poisson fluctuations near a boundary. Further, the renormalized size $s_r$ of the Alexandroff profile decreases from about half to a quarter, which is a characteristic of the square. All three of the square's profiles are distinct from those of the null boundary (Figures~\ref{fig:null_profiles},~\ref{fig:profiles}), leaving no ambiguity over the existence of at least a spacelike boundary. Compared to those of the square, the same profiles for the cylinder (Figure~\ref{fig:sq_v_cyl}(right)) are nearly the same, except for the Alexandroff profile. When the chain length between the past and future spacelike hypersurfaces is spatially independent, likewise there should be no dependence of the Alexandroff profile on spatial position. \par

\subsection{The deformed square}
\begin{figure*}[!pt]
\centering
\includegraphics[width=\linewidth]{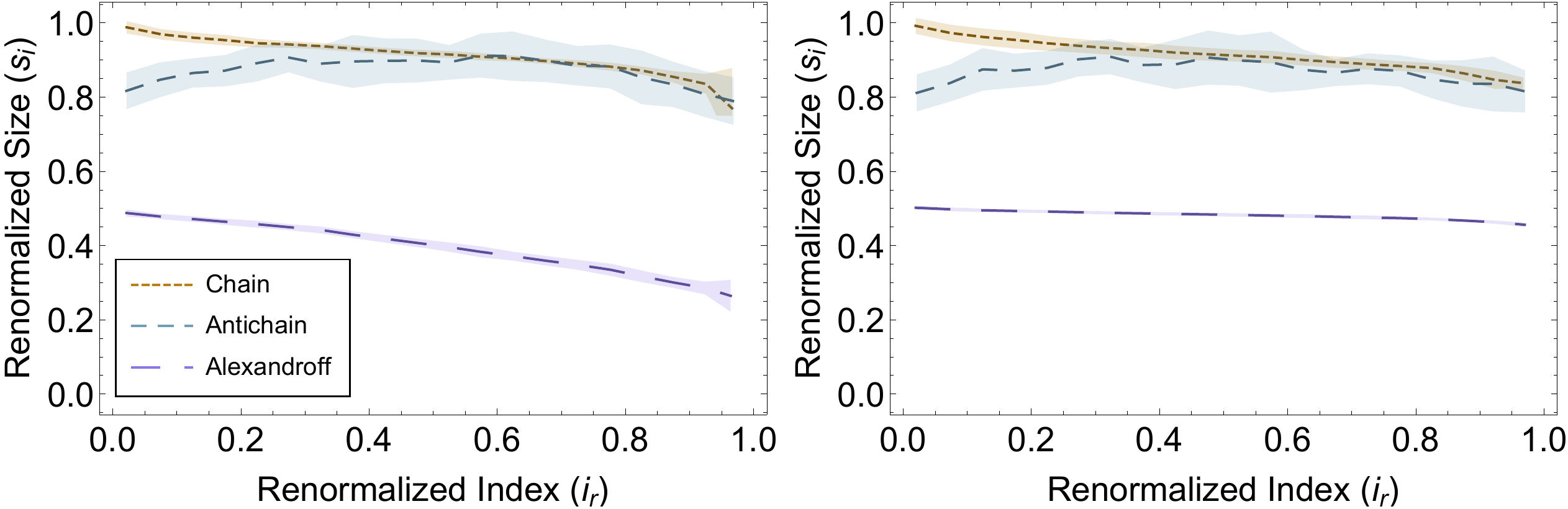}
\caption{{\bf The square versus the cylinder.}}
\label{fig:sq_v_cyl}
\includegraphics[width=\linewidth]{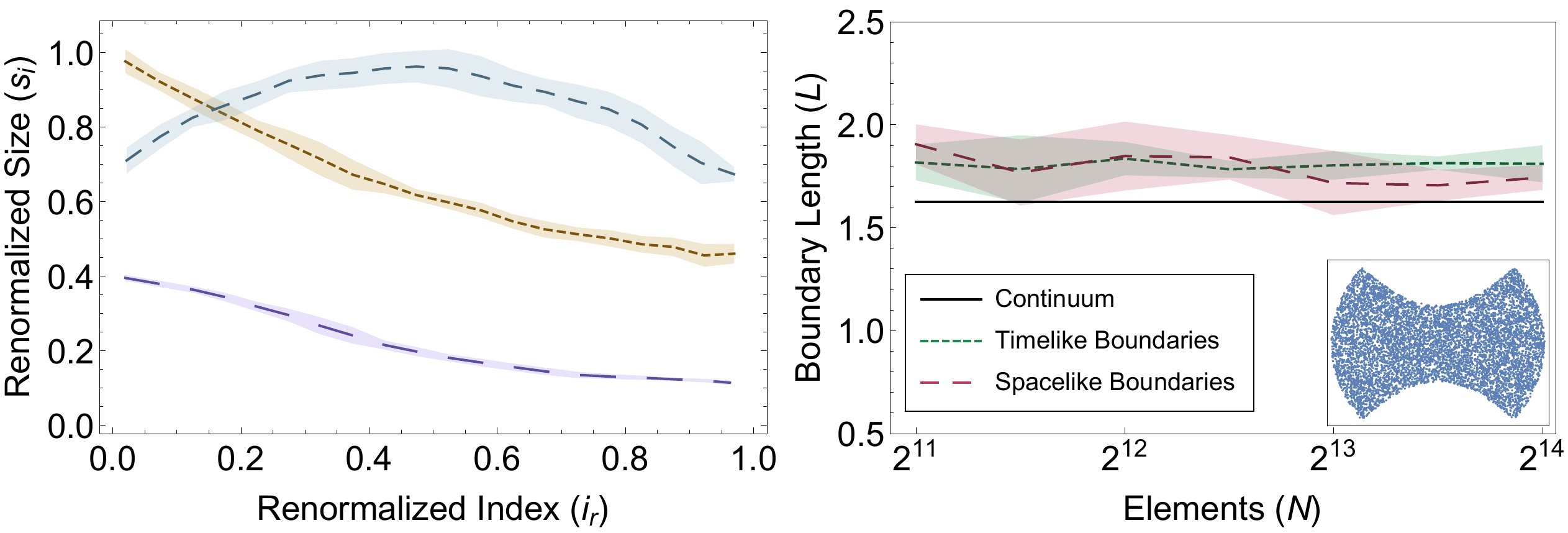}
\caption{{\bf The deformed square.}}
\label{fig:def_sq}
\includegraphics[width=\linewidth]{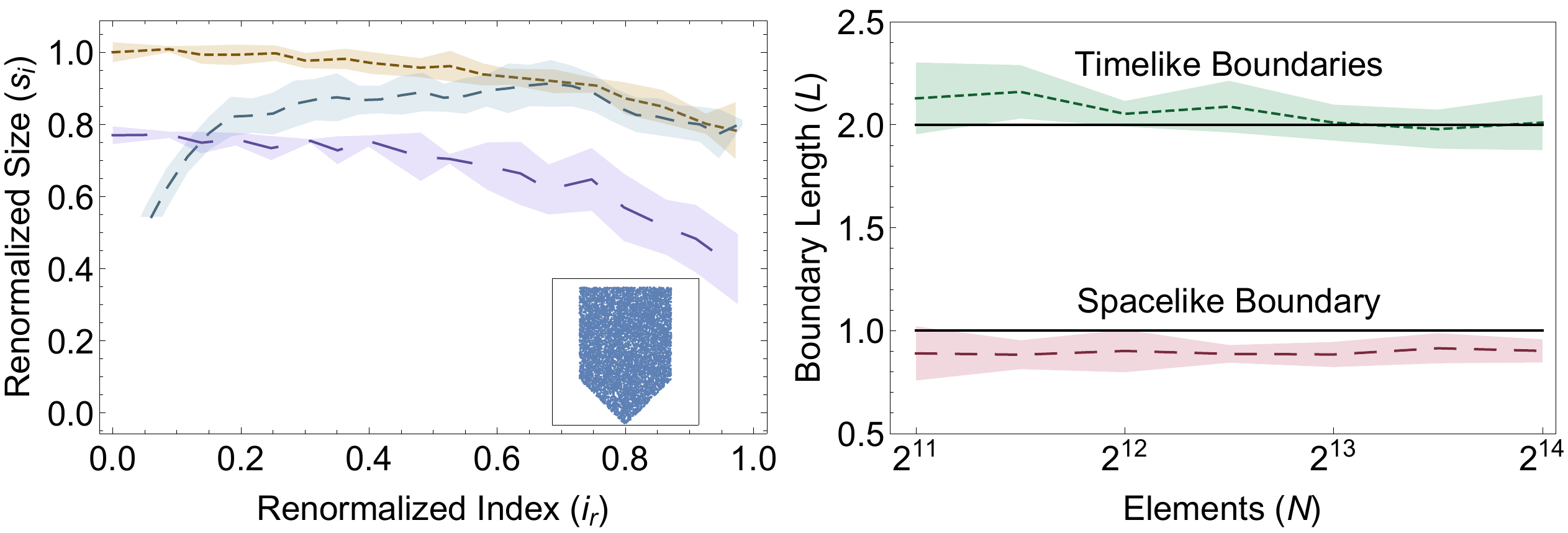}
\caption{{\bf The isosceles right pentagon.}}
\label{fig:pentagon}
\end{figure*}

The second example shows what happens \blue{when} there is a mixture of convex and concave boundaries, shown by the deformed square in the inset of Figure~\ref{fig:def_sq}(right). The left panel of the figure shows the three profiles for this region. The antichain profile indicates the timelike boundary is convex but non-null, since the renormalized size always remains far above zero, i.e., there are no small antichains. The Alexandroff profile differs from the previous example in values but not by behavior, indicating the presence of timelike boundaries and curved spacelike boundaries. \par

The most notable difficulty here is that the longest chain no longer runs through the true spatial origin, but rather through two of the four corners. When a spacelike boundary is concave, the chains in the chain profile are ordered differently, so the method which detects elements near a timelike boundary has some trouble, especially when the extrinsic curvature is large. The monotonically decreasing chain profile could lead one to believe the spacelike boundary is actually convex. Despite this apparent ambiguity, the sign of the spacelike boundary term of the action on expectation gives the sign of the boundary curvature~\cite{buck2015}. \par

The right panel of Figure~\ref{fig:def_sq} shows the results of the timelike and spacelike boundary volume estimation using the methods described in Section~\ref{sec:volume}. All four sides of this region have the same length in the continuum. While the results are close in agreement for the range of causal set sizes shown, they do not yet converge precisely to the continuum limit. It is expected at larger $N$ this convergence occurs, but it is not yet clear what order of magnitude is required. Surprisingly, the value measured for the timelike boundary volume is very close to that for the spacelike boundary volume, despite the fact that the former is an algorithm and the latter an analytic result. One might think it would be worse, since the longest chain is no longer at the spatial origin and some of the assumptions do not hold. For instance, the correlation between radius and number of relations (Figure~\ref{fig:timelike}(left)) is positive rather than negative in this region, so in the first few antichain partitions, candidates are selected close to $r=0$. For this particular choice of width and height, the boundary measurement algorithm still succeeds, but this flaw implies that when the extrinsic curvature of a concave boundary is too large, the algorithm builds a chain directly through the center. Since one may easily test the sign of the boundary curvature, future work will focus on modifications for these cases. \par

\subsection{The pentagon}
In the concluding example, we return to the region shown in Figure~\ref{fig:embedding}(right). For simplicity we take the $(1+1)$-dimensional version, i.e., an isosceles right pentagon with three equal sides, shown by the inset in the left panel of Figure~\ref{fig:pentagon}. The left panel once again shows the three profiles for the region. The chain profile is nearly uniform, indicating the spacelike boundaries are either flat or null and are radially symmetric. The Alexandroff profile is not extremely helpful in this case; it only suggests the existence of timelike boundaries, as it did in the other two examples. The key feature which clarifies the extrinsic geometry of this region is the antichain profile: the curve grows quickly in the first third of the region, and then remains roughly constant in the upper two-thirds. The fact that there are small antichains, along with the shape of the growth, indicates there is a null boundary (see Figure~\ref{fig:null_profiles}). The uniformity in the upper two-thirds strongly suggests that portion of the boundary is flat and timelike. Together with the chain profile, these results also suggest the future spacelike boundary is flat, and this can be confirmed by studying the spacelike boundary term of the action. \par

The right panel of Figure~\ref{fig:pentagon} shows the measurements of the spacelike and timelike boundary volumes. Not surprisingly, the timelike boundary measurement algorithm performs well for flat boundaries, even in the presence of a null boundary. The spacelike boundary volume measurement appears to be consistently below the continuum value, indicating these causal sets are not yet large enough to show convergence.

\section{Conclusion}
\label{sec:conclusion}
By constructing and examining the chain, antichain, and Alexandroff profiles, we have learned how to identify the different types of boundaries of a causal set. We developed a spacetime representation to build maximal chains and antichains as a way to qualitatively describe the causal set. After looking at the profiles for the null boundary in Section~\ref{sec:distributions}, we \blue{could} distinguish a null surface from both a spacelike and timelike boundary, provided it has a small enough extrinsic curvature. The timelike boundary element detection algorithm presented in Algorithm~\ref{alg:candidates} led to a method for the measurement of such a boundary via the guided chain construction in Algorithm~\ref{alg:timelike_measurement}. Finally, we studied the properties of causal sets embedded into three spacetime regions in Section~\ref{sec:examples}. While results here focused on $(1+1)$-dimensional Minkowski spacetime, the techniques can easily be generalized to study higher-dimensional conformally flat spacetimes in future work.

\ack
Thanks to S.\ Surya, \blue{D.\ Krioukov}, P.\ van der Hoorn, and I.\ Voitalov for useful discussions and suggestions. This research was supported by NSF grants CNS-1442999 and CNS-1441828. This research was supported in part by Perimeter Institute for Theoretical Physics. Research at Perimeter Institute is supported by the Government of Canada through the Department of Innovation, Science and Economic Development and by the Province of Ontario through the Ministry of Research and Innovation.

\bibliographystyle{iopart-num}
\bibliography{paper}

\end{document}